# A Data-Driven Policy for Addressing Deployability Issue of FMM FRPs: Resources Qualification and Deliverability

Mohammad Ghaljehei, Mojdeh Khorsand, *Member, IEEE*

*Abstract*— Intensified netload uncertainty and variability led to the concept of a new market product, flexible ramping product (FRP). The main goal of FRP is to enhance the generation dispatch flexibility inside real-time (RT) markets to mitigate energy imbalances due to ramp capability shortage. Generally, the FRP requirements are based on system-wide or proxy requirements, so the effect of FRP awards on the transmission line constraints is not considered. This can lead to FRP deployability issues in RT operation. This paper proposes a new FRP design based on a data-driven policy incorporating ramping response factor sets to address FRP deployability issue. First, a data-mining algorithm is performed to predict the ramp-qualified generators to create the data-driven policy. Then, the FRP awards are assigned to these units while considering effects of post-deployment of FRPs on the transmission line limits. Finally, the proposed data-driven policy is tested against proxy policy through an out-of-sample validation phase that (i) mimics the RT operation of the CAISO, and (ii) represents the expensive ad-hoc actions needed for procuring additional ramping capability to follow realized netload changes. The results show the effectiveness of the proposed data-driven policy from reliability and economic points of view.

## NOMENCLATURE

*Sets and Indices*

| | |
|---|---|
| $g$ | Index of generation resource, $g \in G$. |
| $i$ | Index of solar power generation, $i \in I$. |
| $k$ | Index of transmission line, $k \in K$. |
| $n$ | Index of bus, $n \in N$. |
| $t$ | Index of time interval, $t \in T$. |
| $w$ | Index of training scenario for data-mining algorithm, $w \in W$. |
| $s$ | Index of deployment scenario, $s \in S$. |
| $e$ | Index of block in linearized operating cost of generation resource, $e \in E$. |
| $g(n)$ | Set for generation resources connected to node $n$. |
| $i(n)$ | Set for solar power generations connected to node $n$. |
| $n(g)$ | Set for the node of generation resource $g$. |
| $n(i)$ | Set for the node of solar power generation $i$. |
| $G^F$ | Set for fast-start generation resource. |
| $G^M$ | Set for must-run generation resources in FMM. |

*Parameters and Constants*

| | |
|---|---|
| $c_g^{NL}, c_g^{SD}, c_g^{SU}$ | No-load, shutdown, and startup costs of unit $g$. |
| $P_g^{max}, P_g^{min}$ | Maximum and minimum power capacity outputs of unit $g$. |
| $R_g^{15}$ | 15-min ramp rate of unit $g$. |
| $R_g^{SU}, R_g^{SD}$ | Startup and shutdown ramp rates of unit $g$. |
| $UT_g, DT_g$ | Minimum up and down times of unit $g$. |
| $FRup_t$ | Flexible ramping up requirement in period $t$. |
| $FRdown_t$ | Flexible ramping down requirement in period $t$. |
| $P_k^{max}$ | Thermal rating of transmission line $k$. |
| $PTDF_{nk}$ | Power transfer distribution factor for line $k$ for an injection at bus $n$. |
| $Load_{nt}^{nodal}$ | Nodal load at bus $n$ at period $t$. |
| $\underline{F_g}$ | Minimum operating cost of generation resource $g$ for being committed with minimum power. |
| $B_{ge}$ | Slope of block e in linearized operating cost of unit $g$. |
| $c_g^{UR}, c_g^{DR}$ | Cost of upward and downward FRP of unit $g$. |
| $u_{gt}^h$ | Commitment status of generation resource $g$ at period $t$. |
| $P_{ge}^{max}$ | Maximum power generation of block $e$ in linearized operating cost of generation resource $g$ at period $t$. |
| $Load_t^f$ | Total forecasted load at period $t$. |
| $Load_t^{f,min}$ | Minimum of total forecasted load at period $t$. |
| $Load_t^{f,max}$ | Maximum of total forecasted load at period $t$. |
| $P_{it}^{solar}$ | Power output of solar generation $i$ at period $t$. |
| $P_{it}^{solar,min}$ | Minimum power output of solar generation $i$ at period $t$. |
| $P_{it}^{solar,max}$ | Maximum power output of solar generation $i$ at period $t$. |
| $P_{itw}^{solar,train}$ | Power output of solar generation $i$ at period $t$ at training scenario $w$. |
| $Load_{ntw}^{train}$ | Nodal load at bus $n$ at period $t$ at training scenario $w$. |
| $Load_{ts}^{f,dep}$ | Total load at period $t$ at deployment scenario $s$. |
| $P_{its}^{solar,dep}$ | Power output of solar generation $i$ at period $t$ at deployment scenario $s$. |
| $\zeta_{gts}$ | Ramping response set factor of generation resource $g$ at period $t$ at deployment scenario $s$. |
| $\Delta NL_{ts}$ | Difference of netload of deployment scenario $s$ at period $t+1$ compared to forecasted netload at period $t$. |
| $\overline{ur_{gt}}, \overline{dr_{gt}}$ | Scheduled FMM upward and downward FRP awards of generation resource $g$ in period $t$. |
| $\overline{f_{kts}^{dep,ur}}, \overline{f_{kts}^{dep,dr}}$ | Calculated flow of transmission line $k$ for post-deployment of FRPs at period $t$ at deployment scenario $s$. |

*Variables*

| | |
|---|---|
| $P_{gt}$ | Power generation of generation resource $g$ at period $t$. |
| $P_{nt}^{inj}$ | Net power injection at bus $n$ at period $t$. |
| $u_{gt}, v_{gt}, w_{gt}$ | Variables of unit commitment, startup, and shutdown for generation resource $g$ in period $t$. |
| $ur_{gt}, dr_{gt}$ | Upward and downward FRP awards of generation resource $g$ in period $t$. |
| $P_{gte}$ | Power generation of block $e$ in linearized operating cost of generator $g$ at period $t$. |
| $ur_{gts}^a$ | Upward ramping auxiliary variable of generation resource $g$ at period $t$ at deployment scenario $s$. |
| $ur_{gts}^a$ | Downward ramping auxiliary variable of generator $g$ at period $t$ at deployment scenario $s$. |

## I. INTRODUCTION

### A. Background and Motivation

**O**perational complexities are arising in the modern power systems due to increasing uncertainty and variability in the system netload (i.e., system load minus total renewable power generation and interchange flow). The intensified uncertainty and variability, which are mainly caused by high penetration of the renewable energy resources, lead to a significant need for ramp-down capabilities during sunrise and ramp-up capabilities during sunset, as illustrated by the Duck Curve of the California independent system operator (CAISO) [1]. It is pertinent to note that in the case of insufficient ramp capabilities in the system, the power balance violation can occur, which jeopardizes the reliability and security of the system and cause frequent high penalty prices [2]. To address this issue, some independent system operations, including CAISO and midcontinent independent system operator (MISO), have been implementing flexible ramp products (FRP) in their market. FRP is a market-based product with a ramping capacity held at a time interval to respond to the netload changes (in both directions, up and down) of the next time interval [3]. It is pertinent to note that since the main goal of the FRP is to enhance dispatch flexibility inside the FMMs and RTDs, it is constantly deployed in the fifteen-minute markets (FMMs) and real-time economic dispatch (RTD) markets.

Since the FRP requirements implemented by industry or other work are based on the proxy system-wide or zonal requirements, there is no guarantee that the awarded FRPs will be deployable without violating transmission line limits. The reason can simply be associated with the fact that the FRP post-deployment deliverability within the transmission line limits is disregarded when making decisions on the FRP awards. Ideally, the FRP awards should be assigned to qualified ramp-responsive resources that are not behind the transmission bottlenecks. More advanced techniques to deal with the uncertainty imposed on the power system include, but are not limited to, (i) stochastic programming (e.g., two-stage stochastic models), wherein the uncertainties are explicitly represented and simultaneously solved in the model [4] and (ii) robust optimization, which mitigates worst-case consequences [5]. The independent system operators (ISOs) in the U.S. do not utilize the stochastic programming and the robust optimization in their day-ahead (DA) and real-time (RT) market processes since (i) these approaches have implementation complexity and computational requirements for the real-world scheduling problems [6], and (ii) market pricing and settlement based on these approaches are not well-acceptable by stakeholders and energy market engineers.

Therefore, an approach is desirable in this situation to create a proper balance between decisions efficiency (i.e., deployable FRP awards) and complexity while being practically implementable. In this paper, statistical information and knowledge of the market outcomes under the possible realization of scenarios, data-mining algorithms, and enhanced FRP policies are leveraged efficiently to enhance the decisions on FMM FRP awards, without adding too much disruption to the existing energy market practices and compromising computational efficiency. The final goal is to award the ramp capabilities to the potential generation resources which could deploy them in the RT operation. This goal also has been perused by CAISO to reduce the ad-hoc and out-of-market corrections to create additional ramp capabilities [7] and [8].

### B. Literature Review

Some work [2], [9]–[12] have been seeking to implement FRP in the DA market to increase the ramp-responsiveness and flexibility procurement from available resources. However, the original design of FRP was for the enhancement of the ramp capabilities in RT, and DA FRP is still in its initial design stage for the DA market. At the same time, FRP has been implemented for a couple of years in RT markets of ISOs. References [13]–[22] are mainly associated with incorporating the FRP in FMMs and RTD markets. The FRP for an RTD market was introduced in [13], wherein the ramp capability requirement is set based on the system-wide requirement. A simplistic case study is employed to evaluate FRP markets performance fundamentally. A similar study with the same intentions is conducted in [14]; however, the comparisons are made for a real-time unit commitment (RTUC) market model with 15-min granularity. It is concluded that including FRP in FMM and RT markets enhances overall dispatch flexibility, but stochastic market models have better performance. Reference [15] compares results of a RT market model with FRP constraints (similar to ISO models), i.e., dispatch, prices, settlements, and market efficiency, with a stochastic market model. A single-period market model with FRP is compared with a time-coupled multi-period market model in [16] from efficiency, reliability, and incentive compatibility points of view. Reference [17] discussed the FRP design for the RT economic dispatch in the MISO market, wherein the ramping requirements are set 10-min ahead to manage net load variations and uncertainty. In contrast, the RT dispatch is performed every 5-min. It has been shown that the 10-min ramping capabilities may be depleted in the first 5-min dispatch; thus, the system may not be able to follow the variability and uncertainty of the net load in the second 5-min dispatch. Therefore, a new FRP design is proposed in [17] to address this subtle issue and maintain system reliability and efficiency. The market design based on system-wide FRP policy was improved for the RTUC problems in [18], [19]. A new solar FRP was proposed in [20] for a multi-interval RTD market model, wherein, for estimation of FRP requirement, a Gumbel copula-based joint probability distribution of load and generation forecasting errors was used. However, the RT FRP model presented in [13]–[20] is still based on the system-wide requirements without consideration of which generation resources are more qualified for ramp capability products and/or the effects of transmission lines on the FRP awards deployability.

To address the deliverability issue of FRP, a distributionally-robust multi-interval optimal power flow was proposed in [21] while considering the spatiotemporal correlations among demand uncertainties and wind power. Reference [22] has proposed a robust FRP design for the RTD markets to address the FRP deliverability issues. Also, the corresponding pricing scheme was developed to value the generation resources that provide the FRP. Still, as mentioned earlier, applications of these advanced optimization models for real-world market models are limited due to the scalability and pricing barriers. Reference [23] includes the FRPs constraints into a risk-limiting RT economic dispatch problem, wherein risk has been defined as the "loss-of-load probability." Overall, to address the deployability issue, FRP awards should be given to the generation resources that can deploy them during ramping shortages. The FRP decisions should be made while considering their effects on the transmission line limits.

*C. Contributions and Paper Structure*

In this paper, to address the deployability issue of FRPs, a computationally tractable data-driven policy is proposed for FRPs awards in FMMs, which improves upon the existing industry models. The key idea is to allocate the FRPs to the resources that can effectively deploy their ramping capabilities in the corresponding locations when needed without violating transmission line limits. To this end, the proposed approach uses data-mining algorithms as a mean to generate ramping response factor sets. These ramping response factor sets can be used to determine generation resources ramping responsiveness given during ramping events. Furthermore, the impacts of the post-deployment of FRPs on the transmission line flows are considered; inclusion of post-deployment constraints further stimulates allocation of FRPs to generators that are not located behind transmission bottlenecks. Finally, the performance of the enhanced FMM market model, i.e., the modified model to include the proposed FRP design, is compared against the FMM market model with proxy FRP design through an out-of-sample validation methodology. This validation methodology mimics RT unit commitment (RTUC) of FMM of CAISO.

Fig. 1 illustrates the flowchart of the proposed algorithm to address the deployability issue of FRP in RT operation, wherein, for the proposed approach, an offline process is performed before running the enhanced FMM formulation with the data-driven FRP policy. The data-driven policy generates ramping response factor sets to assign FMM FRP awards to qualified generation resources in the right location with respect to the transmission line limits for post-deployment of FRPs.

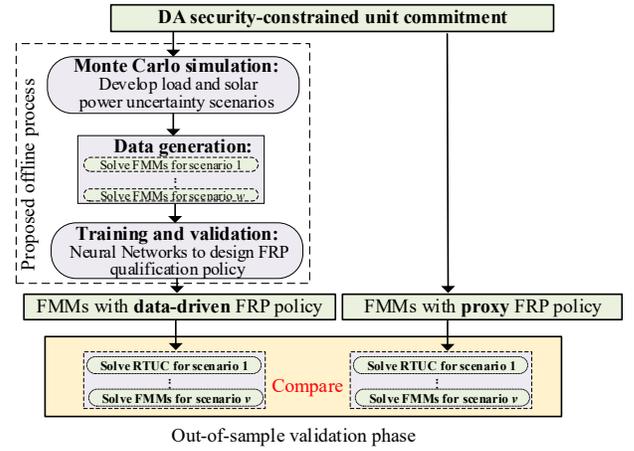

Fig. 1. Comparison of the proxy and data-driven policies for FMM FRP design.

The rest of the paper is organized as follows. Section II presents the different market processes structures, and Section III gives FMMs formulation based on the general FRP design. Section IV focuses on the proposed FRP design, including the data-mining algorithm and the proposed data-driven policy for the FMM FRP design. The out-of-sample validation phase is elaborated in Section V. Simulation results are presented and analyzed in Section VI. Finally, Section VII shows the paper's conclusions.

## II. STRUCTURE OF MARKET PROCESSES

This work focuses on improving the FRP design in the FMMs. However, to get more realistic results and follow the practice in most ISOs, the DA and FMM are run consecutively. The DA market, based on the DA market structure of most ISOs in the US, schedules the energy product to meet the netload while assigning contingency-based reserve and regulation reserve to the generation resources.

The DA market decisions are utilized in the FMMs. The commitment decisions of the must-run (MR) generation resources in the FMMs are fixed based on the DA commitment decisions. At the same time, short-start and fast-start (FS) units can further be committed to satisfying the realized netload for meeting ramping needs.

The FMM structure in this work is based on the FMM structure of CAISO. CAISO performs four RTUC processes, i.e., RTUC#1, RTUC#2, RTUC#3, and RTUC#4, for each trading hour in the CAISO's FMM, spanning from 60 minutes (or four intervals) to 105 minutes (or seven intervals) [24]. These RTUC processes are run on a rolling forward basis, wherein the second 15-min time interval is binding, and the rest are advisory time intervals [24]. This work slightly modifies this FMM structure and performs only a one-process RTUC for each trading hour in the FMM that includes seven intervals with four binning intervals, as depicted in Fig. 2.

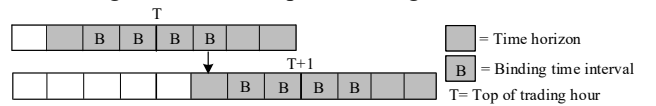

Fig. 2. One-process RTUC run in the validation phase [2].

## III. FMM MODEL WITH GENERAL FRP DESIGN (PROXY POLICY)

The general FMM market model with proxy FRP requirement is given in (1a)-(1ac). The objective function minimizes total operating costs (i.e., linearized variable operating costs plus the fixed costs that include no-load costs, startup costs, and shutdown costs) as follows.

$$minimize \sum_g \sum_t (F_g u_{gt} + \sum_{e=1}^E P_{gte} B_{ge} + c_g^{SU} v_{gt} + c_g^{SD} w_{gt} + c_g^{UR} ur_{gt} + c_g^{DR} dr_{gt}) \quad (1a)$$

The objective function is subject to generation resource operational constraints, transmission network constraints, and general FRP constraints presented through (1b)-(1ac). Constraints (1b) and (1c) link DA and FMM commitment decision variables of the MR and FS generation resources. The generation resources minimum up time and down time constraints are modeled by (1d) and (1e). The power output of generation resources and the limit on each power generation block are given by (1f) and (1g). Constraints (1h) and (1i) enforce 15-min ramp rate limits. Constraint (1j) ensures nodal balance at each bus, while constraint (1k) guarantees the system-wide energy balance between total loads and generation resources throughout the system. Constraint (1l) enforces the transmission line limits. Finally, constraints (1m)-(1q) are associated with modeling the commitment, startup, and shutdown variables of generators.

$$u_{gt} = \overline{u_{gt}^h}, \forall g \in G^M, t \quad (1b)$$
$$u_{gt} \geq \overline{u_{gt}^h}, \forall g \in G^F, t \quad (1c)$$
$$\sum_{s=t-UT_g+1}^{t} v_{gs} \leq u_{gt}, \forall g \in G^F, t \in \{UT_g, \cdots, T\} \quad (1d)$$
$$\sum_{s=t-DT_g+1}^{t} w_{gs} \leq 1 - u_{gt}, \forall g \in G^F, t \in \{DT_g, \cdots, T\} \quad (1e)$$
$$P_{gt} = P_g^{min} u_{gt} + \sum_{e=1}^E P_{gte}, \forall g, t \quad (1f)$$
$$0 \leq P_{gte} \leq P_{ge}^{max} u_{gt}, \forall g, t, e \quad (1g)$$
$$P_{gt} - P_{gt-1} \leq R_g^{15} u_{gt-1} + R_g^{SU} v_{gt}, \forall g, t \geq 2 \quad (1h)$$
$$P_{gt-1} - P_{gt} \leq R_g^{15} u_{gt} + R_g^{SD} w_{gt}, \forall g, t \geq 2 \quad (1i)$$
$$\sum_{g \in g(n)} P_{gt} + \sum_{i \in i(n)} P_{it}^{solar} - Load_{nt}^{nodal} = P_{nt}^{inj}, \forall n, t \quad (1j)$$
$$\sum_n P_{nt}^{inj} = 0, \forall t \quad (1k)$$
$$-P_k^{max} \leq \sum_n P_{nt}^{inj} PTDF_{nk} \leq P_k^{max}, \forall k, t \quad (1l)$$
$$v_{gt} - w_{gt} = u_{gt} - u_{g,t-1}, \forall g, t \quad (1m)$$
$$v_{gt} + w_{gt} \leq 1, \forall g, t \quad (1n)$$
$$0 \leq v_{gt} \leq 1, \forall g, t \quad (1o)$$
$$0 \leq w_{gt} \leq 1, \forall g, t \quad (1p)$$
$$u_{gt} \in \{0,1\}, \forall g, t \quad (1q)$$

Constraints (1r)-(1ac) represent the FMM FRP design based on the proxy policy. The power capacity constraints, including energy power output, with upward and downward FRP awards, are given by (1r) and (1s). Constraints (1t)-(1w) model the limitation on the upward and downward ramp capabilities considering the commitment status of generation resources. The proxy up and down ramping requirements are calculated by (1x) and (1y) and are met through (1z) and (1aa), respectively.

$$p_{gt} + ur_{gt} \leq P_g^{max} u_{gt} + P_g^{max} v_{gt+1}, \forall g, t \in T \quad (1r)$$
$$p_{gt} - dr_{gt} \geq P_g^{min} u_{gt} - R_g^{SD} w_{gt+1}, \forall g, t \in T \quad (1s)$$
$$ur_{gt} \leq R_g^{15} u_{gt} + R_g^{SU} v_{gt+1}, \forall g, t \in \{1, \cdots, T-1\} \quad (1t)$$
$$dr_{gt} \leq R_g^{15} u_{gt+1} + R_g^{SD} w_{gt+1}, \forall g, t \in \{1, \cdots, T-1\} \quad (1u)$$
$$ur_{gt} \leq P_g^{max} u_{gt+1}, \forall g, t \in \{1, \cdots, T-1\} \quad (1v)$$
$$dr_{gt} \leq P_g^{max} u_{gt}, \forall g, t \in \{1, \cdots, T-1\} \quad (1w)$$
$$FRu_t = \max[Load_{t+1}^{f,max} - \sum_i P_{it+1}^{solar,min} - (Load_t^f - \sum_i P_{it}^{solar}), 0], \forall t \in \{1, \cdots, T-1\} \quad (1x)$$
$$FRdown_t = \max[Load_t^f - \sum_i P_{it}^{solar} - (Load_{t+1}^{f,min} - P_{it+1}^{solar,max}), 0], \forall t \in \{1, \cdots, T-1\} \quad (1y)$$
$$\sum_g ur_{gt} \geq FRup_t, \forall t \in \{1, \cdots, T-1\} \quad (1z)$$
$$\sum_g dr_{gt} \geq FRdown_t, \forall t \in \{1, \cdots, T-1\} \quad (1aa)$$

It is pertinent to note that FMM FRP constraints (1r)-(1w) consider the effects of the commitment status on FRP awards in the FMM framework. The FRP design in some prior work is only valid when generation resources are committed in both time intervals $t$ and $t+1$ [2], [9]–[12]. However, such FRP design cannot correctly represent ramping capability from the generation resources committed only during one of the time intervals $t$ or $t+1$. For example, if a generation resource is committed at time interval $t$ and is not committed at time interval $t+1$, it should only be allowed to provide the downward FRP up to its shutdown ramp rate. Also, when a generation resource is not committed at time interval $t$ and is committed at time $t+1$, it should only provide the upward FRP up to its startup ramp rate. The FMM FRP constraints (1r)-(1w) represent the mathematical formulation of these situations for the generation resources that their commitment status varies between two successive time intervals. These formulations can also be extended to the DA framework.

Finally, the dispatch changes of generation resources should be within the FRP awards as presented by (1ab)-(1ac). These constraints are essential as the FRP awards cover both foreseen (variability) and unforeseen (uncertainty) netload changes. Therefore, if a generation resource changes its power generation for covering variability, the associated FRP award should be able to cover it [25].

$$p_{gt+1} - p_{gt} \leq ur_{gt}, \forall g, t \in \{1, \cdots, T-1\} \quad (1ab)$$
$$p_{gt} - p_{gt+1} \leq dr_{gt}, \forall g, t \in \{1, \cdots, T-1\} \quad (1ac)$$

## IV. PROPOSED FRP DESIGN (DATA-DRIVEN POLICY)

In general, a basic way to manage increased uncertainty of net load caused by renewable resources is to procure ancillary services (e.g., reserves and ramp capability products) [15], [26]–[29]. In other words, ISOs model constraints to procure ancillary services such that generators can respond to the deviations of net load. However, since the FRP design proposed by the industry is based on proxy system-wide or zonal requirements, there is no guarantee that these FRPs, which are procured by the market, will be deployable without violating transmission line limits. The reason can simply be associated with the fact that the bus-level deliverability for the post-deployment of FRPs within transmission limits is disregarded when deciding on the FRP awards.

As mentioned in Section I, to cope with the deployability issue of FRPs, it is necessary to assign FRP awards to qualified resources that are not located behind transmission bottlenecks. In this paper, a new framework is proposed to pave the way for achieving the above goal. The proposed FMM framework (i) predicts ramping responses of generation resources in RT, (ii) assigns deployable FRP awards to qualified resources, and (iii) enforces the transmission line constraints for post-deployment of FRP awards. Sections III. A-III. B show the proposed approach to predict the ramping response of generation resources (for generating ramping response factor sets in this paper). Finally, Section III. C presents the proposed data-driven policy for the FMM FRP design considering these ramping response factor sets.

### A. Data Generation for the Data-Mining Algorithm

In this section, the process for generating training data for the data-mining algorithm, which is explained in the next section, is elaborated. First, different load and solar power generation scenarios using the corresponding forecast values and forecast errors are generated through Monte Carlo simulation. Then, FMM presented by (2a)-(2c) is solved for each scenario $w$ to identify how the generators respond to netload changes.

$$\text{minimize} \sum_g \sum_t (F_g u_{gt} + \sum_{e=1}^{E} P_{gte} B_{ge} + c_g^{SU} v_{gt} + c_g^{SD} w_{gt} + c_g^{UR} ur_{gt} + c_g^{DR} dr_{gt}) \quad (2a)$$

subject to:

(1b)-(1i) and (1k)-(1q) (2b)

$$\sum_{g \in g(n)} P_{gt} + \sum_{i \in i(n)} P_{itw}^{solar} - Load_{ntw} = P_{nt}^{inj}, \forall n, t \quad (2c)$$

Please note that the FRP constraints are not incorporated in the FMM formulation (2a)-(2c); the intention of this model is to analyze how different generation resources increase or decrease their output to follow the realized netload changes. After solving the above formulation for each scenario $w$ at each time interval $t$, changes in dispatch setpoints of generators due to their ramping capability provision can be obtained. These data along with net-load, load, nodal solar generations, net-load change, load change, and nodal solar generation changes will form the training dataset for the proposed algorithm, which is explained in the next section. Please note that this stage, i.e., running FMMs based on the Monte Carlo simulation, can also be replaced with available historical data.

### B. Data-Mining Algorithm for Obtaining Ramping Response Factor Sets

The goal here is to assess the deployability of ramping responses for various generation resources based on their locations and allocate FRP effectively. Data-mining algorithms can be utilized to predict ramping response of a generation resource to a given ramping event. This information is then included in the proposed FMM formulation, i.e., (3b)-(3i), to select responsive generators for FRP awards. The machine learning algorithm utilized in this paper is a neural network (NN) regression function. The features are net-load, load, nodal solar power generations, net-load change, load change, and nodal solar power generation changes.

The prediction task should be performed for each time interval $t$ and each generation resource except fast start units as these units can be committed or decommitted in FMM to follow the realized netload. Previous and next time intervals may influence the ramping response of generation resources in the FMM due to its multi-period dispatch characteristic and intertemporal constraints. Therefore, the features explained above are extended to include net-load, load, nodal solar power generations, net-load change, load change, and nodal solar power generation changes of next and previous three intervals in addition to interval $t$ (i.e., seven intervals in total, including time interval $t$). So, the total number of features would be $7 \times (4 + 2 \times |I|)$, where $I$ is the set of the solar power generation.

Please note that both data acquisition and performing the NNs are offline processes before running the actual FMM. Then, the generation resources that are qualified to respond to a set of deployment scenarios can be identified through the concept of the ramping response factor set, which are target values of NN regression functions. Inclusion of the factor sets in FMM are explained in the next section.

### C. FMM Model with Data-Driven FRP Design and Ramping Response Factor Sets

This section presents the FMM formulation incorporating the ramping response factor sets from the proposed data-driven FRP design. First, after obtaining the NN functions from the offline process, close to running the RT FMMs, possible deployment scenarios can be generated based on RT forecasts, including power solar generations and load values. Note that these deployment scenarios can be the most probable scenarios for the load and solar power generations. Then, the trained NN will be utilized to obtain the ramping response factor sets $\zeta_{gts}$.

Second, the proposed data-driven FRP design is extended based on the existing FRP formulation (1r)-(1ac). In this new formulation, the qualified generation resources capable of deploying FRP awards concerning the deployable scenario set $S$, are required to satisfy the system-wide FRP requirements. At the same time, the transmission line flow limits are enforced for post-deployment of ramping capabilities.

For each time interval $t$, the deployment scenarios can be divided into two *upward* and *downward* deployment scenarios based on the value of $\Delta NL_{ts}$, defined by (3a).

$$\Delta NL_{ts} = Load_{t+1s}^{f,dep} - Load_t^f + \sum_i P_{it+1s}^{solar,dep} - \sum_i P_{it}^{solar}, \quad s, t \in \{1, \cdots, T-1\} \quad (3a)$$

If $\Delta L_{ts}$ is greater than zero for time interval $t$, then scenario $s$ is considered upward deployment scenarios with the positive ramping requirement at time interval $t$. Otherwise, it is downward deployment scenarios at time interval $t$. Constraints (3b)-(3e) present the proposed data-driven policy for upward FRP, wherein $\Delta NL_{ts} > 0$ for time interval $t$ and deployment scenario $s$. The enhanced formulation utilizes positive auxiliary variables, i.e., $ur_{gts}^a$, to represent the upward ramping response of generation resources at the deployment scenarios. Constraint (3b) assigns ramp capabilities to the qualified generation resources by using the ramping response set factors $\zeta_{gts}^{fru}$. The upward FRP award is set to be greater than all the upward auxiliary variables through (3c). Also, summation over the auxiliary variables should meet the ramping requirement of the deployment scenario, i.e., $\Delta L_{ts}$, by (3d). Finally, constraint (3e) models transmission line constraint for upward FRP post-deployment for the ramping scenarios. In constraint (3e), the first term is the pre-activation flow of transmission line, the second term is the change in the flow due to the upward FRP activation, the third term is related to the change in flow due to solar generation change, and finally, the fourth term is associated to the change in flow due to load change.

$ur_{gts}^a \geq \zeta_{gts} R_g^{15}, g \in G^M, s, t \in \{1, \cdots, T-1\}$ (3b)

$ur_{gts}^a \leq ur_{gt}, g, s, t \in \{1, \cdots, T-1\}$ (3c)

$\sum_{g \in G} ur_{gts}^a \geq \Delta NL_{ts}, \forall s, t \in \{1, \cdots, T-1\}$ (3d)

$-P_k^{max} \leq \sum_n P_{nt}^{inj} PTDF_{nk} + \sum_g ur_{gts}^a PTDF_{n(g)k} + \sum_i (P_{it+1s}^{solar,dep} - P_{it}^{solar}) PTDF_{n(i)k} - \sum_n (Load_{nt+1s}^{nodal,dep} - Load_{nt}^{nodal}) PTDF_{nk} \leq P_k^{max}, k, s, t \in \{1, \cdots, T-1\}$ (3e)

The symmetric formulation can be presented for the enhanced data-driven policy for downward FRP, given in (3f)-(3i).

$dr_{gts}^a \geq -\zeta_{gts} R_g^{15}, g \in G^M, s, t \in \{1, \cdots, T-1\}$ (3f)

$dr_{gts}^a \leq dr_{gt}, g, s, t \in \{1, \cdots, T-1\}$ (3g)

$\sum_{g \in G} dr_{gts}^a \geq -\Delta NL_{ts}, \forall s, t \in \{1, \cdots, T-1\}$ (3h)

$-P_k^{max} \leq \sum_n P_{nt}^{inj} PTDF_{nk} - \sum_g dr_{gts}^a PTDF_{n(g)k} + \sum_i (P_{it+1s}^{solar,dep} - P_{it}^{solar}) PTDF_{n(i)k} - \sum_n (Load_{nt+1s}^{nodal,dep} - Load_{nt}^{nodal}) PTDF_{nk} \leq P_k^{max}, k, s, t \in \{1, \cdots, T-1\}$ (3i)

The proposed FMM with the enhanced data-driven FRP policy is given in (3g)-(3k).

$minimize \sum_g \sum_t (F_g u_{gt} + \sum_{e=1}^E P_{gte} B_{ge} + c_g^{SU} v_{gt} + c_g^{SD} w_{gt} + c_g^{UR} ur_{gt} + c_g^{DR} dr_{gt})$ (3g)

subject to:
(1b)-(1ac) and (3b)-(3i) (3k)

Please note that the transmission constraints presented by (3e) and (3i) add $T \times K \times S$ constraints to the FMM formulation. However, most of these FRPs post-deployment transmission line constraints can be superfluous and do not set up the feasibility space of the FMM problem. In this paper, an iterative procedure is presented to avoid the computational burden of the proposed approach by considering only binding FRPs post-deployment transmission line constraints, wherein the minimum set of these constraints that limits the feasible region is set up. More specifically, the transmission line flow cuts associated with the FRPs post-deployment, added to the FMM problem using the iterative procedure (similar to the branch and cut procedure), serve as umbrella constraints for all the transmission line constraints for FRPs post-deployment. To do so, first, the FMM problem (3g)-(3k) is solved without considering constraints (3e) and (3i). Then, transmission line flows for the post-deployment of FRPs are calculated through equations (3l) and (3m). Suppose calculated flows are out of the limit of the transmission lines. In that case, the corresponding transmission line constraints for post-deployment of FRPs are added to the FMM problem. This process continues until no other line can be found to be problematic during post-deployment of FRPs.

$\overline{f_{kst}^{dep,ur}} = \sum_n P_{nt}^{inj} PTDF_{nk} + \sum_g ur_{gts}^a PTDF_{n(g)k} + \sum_i (P_{it+1s}^{solar,dep} - P_{it}^{solar}) PTDF_{n(i)k} - \sum_n (Load_{nt+1s}^{nodal,dep} - Load_{nt}^{nodal}) PTDF_{nk}, k, s, t \in \{1, \cdots, T-1\}$ (3l)

$\overline{f_{kst}^{dep,dr}} = \sum_n P_{nt}^{inj} PTDF_{nk} - \sum_g dr_{gts}^a PTDF_{n(g)k} + \sum_i (P_{it+1s}^{solar,dep} - P_{it}^{solar}) PTDF_{n(i)k} - \sum_n (Load_{nt+1s}^{nodal,dep} - Load_{nt}^{nodal}) PTDF_{nk}, k, s, t \in \{1, \cdots, T-1\}$ (3m)

## V. OUT-OF-SAMPLE VALIDATION PHASE

It is essential to have a proper validation phase that mimics the RT operation to compare the effectiveness of different FRP policies from reliability and economic aspects. Therefore, this paper presents an out-of-sample validation phase based on the FMM of CAISO, which includes out-of-sample 15-min netload scenarios. While running the FMMs under the out-of-sample scenarios, dispatch change of MR generation resources between two successive time intervals is limited by their corresponding upward and downward FRP awards as given in (4a) and (4b), respectively.

$P_{gt} - P_{gt-1} \leq \overline{ur_{gt}} u_{gt-1} + R_g^{SU} v_{gt}, \forall g, t \geq 2$ (4a)

$P_{gt-1} - P_{gt} \leq \overline{dr_{gt}} u_{gt} + R_g^{SD} w_{gt}, \forall g, t \geq 2$ (4b)

It is pertinent to note that if there is insufficient system ramping capability to follow the netload changes, power balance violation can happen in the validation phase considering the VOLL, and the FS generation resources can still be committed. These additional adjustments and commitments can be translated into the ad-hoc and out-of-merit operator actions, which can be potentially expensive for procuring additional ramp capabilities to meet changes in the netload. The reason can be associated with the fact that operators do not constantly adjust these ad-hoc actions to make a trade-off between the additional ramp needed and the RT operating costs.

## VI. NUMERICAL STUDIES AND DISCUSSION

A modified IEEE 118-bus system is utilized for performing simulation analyses. In addition, CPLEX v12.8 is utilized to solve the different DA and RT market model processes on a computer with an Intel Core i7 CPU @ 2.20 GHz, 16 GB RAM, and a 64-bit operating system.

### A. Assumptions and System Data

The modified 118-bus IEEE test system has 51 generation resources, three solar power generations (located at buses 25, 55, and 89 with 20 %, 20 %, and 60 % share of total solar power generation, respectively), 91 loads, and 186 transmission lines [30]. This paper only includes solar power generation in the simulation due to the high ramping needs it imposes on the electric system. Still, the proposed model can be easily be extended to include any variable renewable resource. Hourly and 15-min load and solar power generation profiles of two days, i.e., January 15, 2020, and September 2, 2021, were chosen from the CAIOS's actual data [31]. The uncertainty of hourly load and solar power generation for the DA model is considered to be ~5%, based on which the associated FMM uncertainty can be calculated through the total probability theory [17] and [32], i.e., $\sigma_{hourly} = 2\sigma_{15-min}$. Also, 1.96 standard deviations that is equivalent to 95% confidence level is considered for system-wide FRP requirement calculation. Power balance violation can occur in different market processes in the case of insufficient ramp capability at the cost of 10000 $/MW. In the FMMs, 21 generation resource with the maximum capacity of 50 MW are considered as FS units, i.e., $G^F$. The NN functions were trained through the Python v.3.7 with Scikit-learn library. 5000 scenarios were generated in the data-generations phase, 75% and 25% of which were respectively used as training and testing datasets. The NN functions have 3 hidden layers with 100, 100, and 25 neurons. The number of features of NNs with three bulk solar power generation units is $7 \times (4 + 2 \times 3) = 70$.

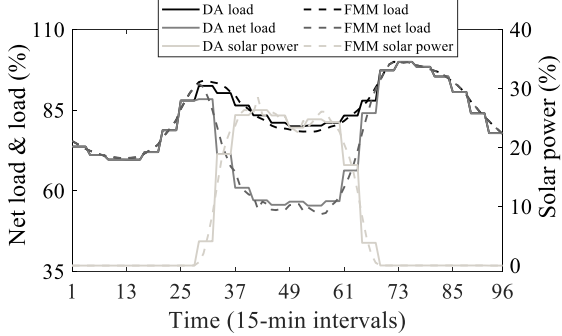

Fig. 3. Hourly and 15-min load, netload, and solar power profiles for test day one.

### B. Simulation Results

For the first test day, profiles are shown in Fig. 3; Fig. 4 compares the total number of FS units commitment reduction against the total violation improvement in the validation phase for the out-of-sample load and solar power generations scenarios. As Fig. 4 illustrates, most of the results of the out-of-sample scenarios are located in the first quadrants of the coordinate plane, in which the data-driven policy is effectively capable of lowering the total violation and the total number of FS units commitment in RT compared to the proxy policy. This improvement is because the data-driven policy preemptively assigns the FRP awards to the ramp-qualified units capable of delivering their products concerning the transmission line limits. It is pertinent to note that a lower need for committing FS units indicates less necessity for ad-hoc or out-of-merit expensive adjustments by ISOs in the RT operation. This goal also is being perused by the CAISO through the FRP nodal delivery test [8]. These results show the performance of the data-driven policy for the FMM FRP design in awarding the FRPs to the generation resources and locations that can be dispatched for following the realized load and solar power generation changes.

Table I lists the number of out-of-sample scenarios in which the data-driven policy outperforms the proxy policy with respect to the economic need for ad-hoc corrections and reliability metrics. Based on this table, it can be seen that the proposed FRP design leads to 100%, 68%, and 97.4% of out-of-sample scenarios having less real-time operating costs (the violation cost with VOLL is removed), total violation, and commitment of FS units, respectively. Comparisons with respect to the real-time operating costs, while the violation cost with VOLL is excluded, are essential for avoiding subjective analyses as the results are sensitive to the value of VOLL.

TABLE I. NUMBER OF OUT-OF-SAMPLE SCENARIOS WITH IMPROVEMENT IN RT OPERATION, TOTAL NUMBER= 500, FIRST TEST DAY

| Metric | |
|---|---|
| # Scenarios with RT cost (excluding violation cost) improvement | 500 |
| # Scenarios with reduction in number of total commitments of FS units | 487 |
| # Scenarios with total violation improvement | 340 |

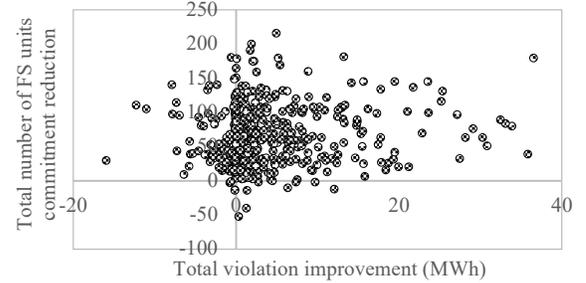

Fig. 4. Total violation improvement versus the total number of FS units commitment reduction for the proposed policy compared to the proxy policy (first test day).

In this paper, to be more objective, two metrics, i.e., the real-time operating costs excluding the violation cost and the total violations, are compared simultaneously. To do so, three statistical reliability measures are presented in Table II to compare and evaluate the security extent in the RT operation due to implementing two proxy and data-driven approaches. Also, Fig. 5 illustrates improvements of the RT operating costs, excluding the violation cost for each 15-min time interval in RT operation. For each time interval, the points on the top and bottom represent the first and third quartiles of improvement, while the height of the bar shows the median average improvement. The results given in Table II show that the data-driven policy outperforms the proxy policy with respect to reliability statistical measures. It is pertinent to note that reducing maximum violation is important when the operator is concerned with the worst-case violation. Also, it can be seen in Fig. 5, the proposed approach leads to RT operating cost (excluding violation cost) improvement in almost all the time intervals through enhancing the FRP decisions. In other words, based on Table II and Fig. 5, while the data-driven policy meets more load, the RT operating costs (not including the violation cost) of the data-driven policy are lower than those of the proxy policy.

The FMM and RT operating costs are tabulated in Table III for proxy and data-driven policies. The data-driven policy leads to relatively higher FMM operating costs (about 0.35% increase) in contrast to the other policy; however, it results in much less ad-hoc expensive adjustments in the RT operation by having less RT operating cost. The average RT operating cost of the proxy policy over all of the out-of-sample scenarios equals $ 2313k, which is significantly greater than the one associated with the data-driven policy, which is $ 2159k. The metrics associated with the number of total commitments of FS units are presented in Table IV. These additional commitments can be translated into the ad-hoc and out-of-merit operator actions, which is likely a high-cost procedure for procuring additional ramp capabilities to meet changes in the netload. The reason can simply be associated with the fact that operators do not constantly tune these ad-hoc actions to make a trade-off between the additional ramp needed and the RT operating costs. According to Table IV, it can be observed that the data-driven policy efficiently lowers the necessity of committing additional expensive FS units.

TABLE II. COMPARISON OF METRICS ASSOCIATED WITH TOTAL VIOLATION IN RT OPERATION (FIRST DAY)

| Metric | Proxy policy | Proposed policy |
|---|---|---|
| Average [Total violation] (MWh) | 28 | 24 |
| Σ Total violation (MWh) | 13852 | 12023 |
| Max Total violation (MWh) | 140 | 107 |

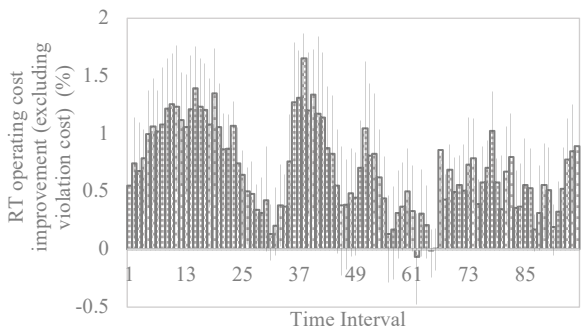

Fig. 5. Total violation improvement versus the total number of FS units commitment reduction for the proposed policy compared to the proxy policy.

TABLE III. COMPARISON OF METRICS ASSOCIATED WITH OPERATING COSTS FOR FMMs AND RT OPERATION (FIRST DAY)

| Metric | Proxy policy | Data-driven policy |
|---|---|---|
| FMM operating costs (K$) | 1171 | 1175 |
| RT operating costs | | |
| Ave (K$) | 2313 | 2159 |
| Max (K$) | 6805 | 5500 |

TABLE IV. COMPARISON OF METRICS ASSOCIATED WITH THE NUMBER OF TOTAL COMMITMENTS OF FS UNITS IN RT OPERATION (FIRST DAY)

| Metric | Proxy policy | Data-driven policy |
|---|---|---|
| Σ Number of total commitments of FS units | 198803 | 166163 |
| Ave [Number of total commitments of FS units] | 398 | 332 |
| Max Number of total commitments of FS units | 747 | 666 |

Each FMM that includes seven intervals for each trading hour takes around 1.05 seconds for the proxy policy to be solved, while the one for the data-driven policy takes 5.6 seconds. Also, the NN functions need around ~6-25 seconds to be trained. Since the NN functions are independent, please note that multiple machines can be used to train them in parallel. To further evaluate the performance of the proposed data-driven policy, one additional test day (i.e., September 2, 2021) was chosen from CAISO's data profiles, as illustrated in Fig. 6. The associated results over the out-of-sample scenarios are given in Tables V-VI and Fig. 7. Based on these results, it can be observed that the proposed data-driven policy outperforms the proxy policy from economic, need for additional ad-hoc actions, and reliability points of view.

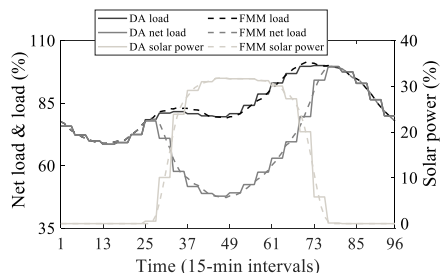

Fig. 6. Hourly and 15-min load, netload, and solar power profiles for the second test day.

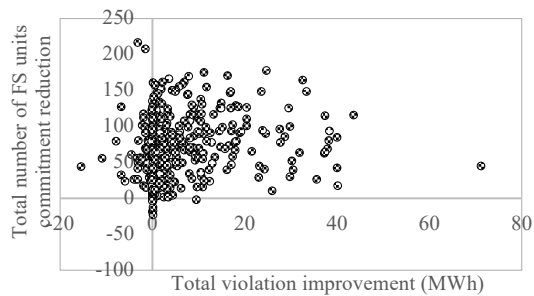

Fig. 7. Total violation improvement versus the total number of FS units commitment reduction for the proposed policy compared to the proxy policy (second test day).

TABLE V. NUMBER OF OUT-OF-SAMPLE SCENARIOS WITH IMPROVEMENT IN RT OPERATION, TOTAL NUMBER= 500, SECOND TEST DAY

| Metric | |
|---|---|
| # Scenarios with RT cost (excluding violation cost) improvement | 498 |
| # Scenarios with reduction in number of total commitments of FS units | 347 |
| # Scenarios with total violation improvement | 488 |

TABLE VI. RESULTS FOR FMMs AND RT OPERATION (SECOND TEST DAY)

| Metric | Proxy policy | Data-driven policy |
|---|---|---|
| FMM operating costs (K$) | 1092.0 | 1094.7 |
| RT operating costs | | |
| Ave (K$) | 1876 | 1689 |
| Max (K$) | 5160 | 4146 |
| Total violation in RT operation | | |
| Average [Total violation] (MWh) | 19 | 14 |
| Σ Total violation (MWh) | 9476 | 7216 |
| Max Total violation (MWh) | 101 | 76 |
| Number of total commitments of FS units in RT operation | | |
| Σ Number of total commitments of FS units | 172923 | 140386 |
| Ave [Number of total commitments of FS units] | 346 | 281 |
| Max Number of total commitments of FS units | 830 | 703 |

VII. CONCLUSION

The general FRP design implemented in most ISOs utilizes a proxy ramping requirement for coping with the ramp capability shortage caused by high variability and uncertainty in the netload. However, these requirements do not consider the transmission line limits' influence when making decisions on FRP awards. This ignorance can cause a significant problem, as also stated by CAISO [7], since the deployability of generation resources FRP awards are dependent not only on its operational limits but also on the ability of the resource to deliver this product in RT operation with respect to the transmission network limits. Therefore, the ISO should carefully consider which generation resources are qualified for awarding the FRPs while considering the limitation of transmission networks during the decision-making on FRP. In this paper, the concept of data-driven FRP policy is introduced and proposed to improve the FRP awards' deployability. To this end, first, by utilizing the NN algorithm, the generation resources ramping response to a set of deployment scenarios are predicted. Then, the system-wide ramping requirement is firstly met with these eligible resources, and at the same time, the transmission line limits are enforced for post-deployment of FRP awards. To evaluate the performance of different FRP policies, an out-of-sample validation phase is presented, which mimics the RT operation of the CAISO's market and provides valuable insights about the ad-hoc actions needed to increase the ramp capability to follow the realized netload. Based on the load and solar power generation profiles of CAISO, the obtained results show that the enhanced FMM FRP design improves the deployability of FRPs with minimal disruption to existing RT market models. Furthermore, the proposed data-driven policy for FRP designs leads to (i) less RT operating costs, (ii) less potential violation in RT operation, (iii) less need for expensive committing FS units in RT operations, and less need for ad-hoc corrections.